 \magnification=\magstep1 
 \font\romnine=cmr9 scaled\magstep0
 \font\medbigrom=cmr10 scaled\magstep1
 \def\romtwelve{\medbigrom}
 \def\Tr{\mathop{\rm Tr}\nolimits} 
 \def\lsheader#1{{
  \removelastskip\vskip 20pt plus 40pt \penalty-200 \vskip 0pt plus -32pt
  \NI\bf #1}\nobreak\medskip\nobreak}
 \def\standardpage{\hsize=6 true in \vsize= 8.5 true in \hoffset=0.25 true in}
 \def\tenstart{\normalbase \tenrm ~~~~~~} 
 \overfullrule=0pt
 \def\refmark{} \def\eq#1{Eq.\ (#1)}  \def\eqs{Eqs.\ }
 \def\NI{\noindent}
 \def\pritem#1{\vskip .04in\item{[#1]}}
 \def\upref#1{$^{#1}$}
 \newcount\ftnum  \newcount\checknum  \newif\ifeqnerr
  \def\aq#1{\global\advance\ftnum by 1 \xdef#1{\the\ftnum}}
  \def\neweq#1{{\global\advance\checknum by 1
     \edef\chqtmp{\the\checknum}
     \checkeq#1}}
 \def\checkeq#1{\ifx#1\chqtmp\global\eqnerrfalse
     \else \global\eqnerrtrue
         \message{ALLOCATION ERROR for \string#1:
            preassigned #1, in sequence \chqtmp.}\fi}

 \def\aqx#1#2{\xdef#2{\the\ftnum#1}}
 \def\neweqx#1#2{{\edef\chqtmp{\the\checknum#1}
     \checkeq#2}}

 \def\normalbase{\baselineskip 12pt}\standardpage
 \footline={\ifnum\pageno=1 \hfill \else\hss\tenrm\folio\hss\fi}
 \tenstart

 \aq\schrodinger \aq\mann \aq\destruct \aq\garg \aq\chuang \aq\timebook
\aq\bounds

 \centerline{\romtwelve MEASURE OF ENTANGLEMENT}
 \bigskip
 \centerline {L. S. Schulman and D. Mozyrsky}
 \medskip
 \centerline {\romnine
               Physics Department, Clarkson University}
 \centerline {\romnine 
               Potsdam, NY 13699-5820 USA}
 \bigskip
 
{\narrower      \romnine \baselineskip 10pt
  \centerline{ABSTRACT} \vskip 2pt
 The extent to which a given wave function, $\psi$, is entangled is measured by
minimizing the norm of $\psi$ minus all possible unentangled functions. This
measure is given by the largest eigenvalue of $\psi^\dagger \psi$, considered
as an operator. The definition is basis independent.
 \par}

\bigskip

\NI
PACS Numbers:
 03.65.Bz, 
 89.80.+h, 
 02.30.Wd  

 \lsheader{~}

   Entanglement was considered by Schr\"odinger to be the essence of quantum
behavior.\upref{\schrodinger\refmark} It is the central concept in the EPR
paradox and Bell inequalities,\upref{\mann\refmark} and now---because of its
effective destruction of interference\upref{\destruct\refmark}---looms as a
significant factor\upref{\garg\refmark} in the construction of systems intended
to retain quantum coherence.\upref{\chuang,\timebook\refmark} For such systems
especially, the distinction between {\it decoherence} and {\it error} is
important. Decohered pieces of wave function---pieces that have
become entangled with environmental degrees of freedom---can {\it
never} provide the useful interference on which the quantum computer
depends.\upref{\destruct\refmark} Moreover, when one combines the effect of
several small entanglement episodes, it is in the {\it amplitude} that the
damage occurs, meaning that this is particularly effective at the destruction
of interference.  Error, on the other hand, leaves one within the original
Hilbert space. This allows corrective measures to be applied more easily. Also,
under some circumstances, if the errors occur within a small number of degrees
of freedom (say a phase), $N$ such errors will only cause a $\sqrt{N}$ defect.
It is thus important to have sharp criteria for distinguishing decoherence from
error. Our measure of entanglement provides such a criterion.

   If $\psi(x,y)$ is a function of two variables, $x$ and $y$, it is said to be
entangled if it cannot be written as a product of two separate functions of $x$
and $y$. It is clear that some functions are ``more" entangled than others:
there should be a sense in which the state $(|+-\rangle -|-+\rangle)/\sqrt{2}$
is highly entangled, while a spin in a quantum computer that has weakly
interacted with a phonon is only slightly entangled (and therefore may continue
to function successfully in the computer). We next provide a precise
measure of entanglement.  The mathematical basis of our definition is
straightforward, so that we will only give general indications of proofs.

   Having defined our measure of entanglement we also apply it to two extreme
situations: a characterization of maximally entangled states and a criterion
for the analysis of only slightly entangled states. For the latter case, we
mention results on the possible entanglement of a confined system with its
walls, in which our criterion allows a precise measure of the damage done
(which interestingly may sometimes be zero).

   Two criteria for the degree of entanglement come to mind. Let
 $$J \equiv \min_{\rho,\sigma}\int dx dy |\psi(x,y)-\rho(x)\sigma(y)|^2
 \eqno(1)$$
 and
 $$ K\equiv \max_{f,g}
       \left|\int dx dy \psi^*(x,y)f(x)g(y)\right|  \eqno(2)$$
 with, in the second definition, normalized $f$ and $g$ (but not in the first
definition). In fact, both definitions produce essentially the same functions,
as we show below. Our measure of entanglement will be $J$. Clearly a function
that is not entangled will have $J=0$. It also turns out that $J=1-K^2$.

   Note that this definition is independent of basis and only depends on the
choice of coordinate system for the configuration space of the degrees of
freedom.

   The mathematical result is most easily stated in matrix form. Let $A$ be an
arbitrary $n\times n$ matrix and as above (except for notational changes)
 $$J \equiv \min_{u,v}\sum_{ij} |A_{ij}-u_i v^*_j|^2    \eqno(3)$$
 Then if $u$ and $v$ minimize $J$, they satisfy the following
 $$  A^\dagger u = v\|u\|^2 \,, \quad    Av=u\|v\|^2         \eqno(4)$$
 from which follow
 $$ AA^\dagger u=\lambda u \,,\quad  A^\dagger A v=\lambda v \eqno(5)$$
 and
 $$ \lambda =\|u\|^2\|v\|^2 = \hbox{~maximum eigenvalue of~}A^\dagger A
  \eqno(6)$$
 Defining $\tilde u=u/\|u\|$, $\tilde v=v/\|v\|$, and $S_2(B) \equiv \Tr
B^\dagger B$ for a matrix $B$, we can write $J= S_2\left(A - \sqrt{\lambda}
|\tilde u\rangle \langle\tilde v| \right)$. It is also clear that
$\langle\tilde u |A| \tilde v\rangle =\sqrt{\lambda}$, which is real, and
finally $J=1-\lambda$.

   Once you know that you want to prove the above assertions, verification is
straightforward. (It is a simple variational problem.) The vector, $v$, was
introduced as a complex conjugate to make the matrix form more transparent.
However, because the original object, $\psi(x,y)$, is a Hilbert space vector,
rather than operator, this may not always be the best notation.

   We next show that the functions $\tilde u$ and $\tilde v$ that minimize $J$
also maximize $K$. For arbitrary, normalized, $w$ and $x$, let $\widetilde J
\equiv S_2(A-\gamma |w\rangle \langle x|)$, where $\gamma$ is an arbitrary
complex constant. If we adjust $|x\rangle$ to make $\langle x|A|w\rangle$
real, then 
 $$\widetilde J = S_2(A) + |\gamma|^2 - \langle x|A|w\rangle (\gamma +\gamma^*)
\eqno(7)$$
 Taking $\gamma$ to be real obviously can only reduce $J$. Since \eq{7} holds
for any real $\gamma$, it is seen that maximizing $ \langle x|A|w\rangle $ is
the same as minimizing $J$.  It then follows that $\lambda=\gamma^2$, etc.

   Using the measure of this paper, the state $(|+-\rangle - |-+\rangle)/
\sqrt{2}$ gives rise to a matrix $A=i\sigma_y/\sqrt2$, and the maximal
eigenvalue of $A^\dagger A$ is $1/2$. This is the maximal degree of
entanglement. Actually it is easy to see that for $n$-state systems wave
functions having the matrix $A_{jk}= \delta(k,\pi_j)\exp(i\phi_{jk})/\sqrt{n}$
are maximally entangled, where $\pi_j$ is any permutation on $n$ objects and
$\phi_{jk}$ is any real matrix. The corresponding eigenvalue is $1/n$, so that
$J=1-1/n$. To see that this is maximal, note that for any $A$ all eigenvalues
of $A^\dagger A$ must be real and non-negative ($\langle \phi| A^\dagger A
|\phi\rangle  = \|A\phi\|^2$). Hence minimizing the maximal eigenvalue is
achieved by having all eigenvalues equal. Moreover, for $A$ arising from a
normalized wave function, $S_2(A)=1=$ sum of the eigenvalues.

   In another article,\upref{\bounds\refmark} the entanglement measure defined
here will be applied to a state that is at worst only slightly entangled.  (The
entanglement there arises from the inevitable interaction of a confined
system---the putative quantum computer---with the ``walls" keeping it in
place.) Using the operator-formalism test developed here, it is shown that in
fact {\it no} decoherence need arise, despite an apparent intertwining of
coordinates. An advantage of our present technique is that it is not necessary
to find the optimum states explicitly---only to solve the eigenvalue problem.
The same is true when there is entanglement: the operator method developed here
measures the entanglement, without requiring knowledge of the optimal factored
state (although in that application enough is known of the operators to produce
the eigenfunctions, should that be desired).  We remark that for this
application the index of ``$A$" is continuous, but that this fact presents no
complication.

   The generalization of the entanglement criterion to three or more degrees of
freedom is straightforward.  From the minimization problem for 
 $$J_3 \equiv \min \int |\psi(x,y,z)-\gamma e(x) f(y) g(z)|^2 dxdydz$$
 (with $e$, $f$ and $g$ normalized) one obtains 
 $$\gamma e(x)^* =\int \psi(x,y,z)^* f(y)g(z)dydz  \eqno(8)$$
 and cyclic permutations of $(e,f,g)$, with $1-\gamma^2$ again the minimum of
$J_3$ (and $\gamma$ real). \eq{8} and its permutations (and its obvious
$N$-particle generalizations) form a non-linear mapping among the Hilbert
subspaces. Although a simple solution as in \eqs(4--6) is not
available, iterative mapping techniques can be applied.

 \lsheader{Acknowledgements}
  We thank V. Privman for helpful discussions. This work was supported in part
by the United States National Science Foundation grant PHY 93 16681.

 \lsheader{References}

 \pritem{\schrodinger} E. Schr\"odinger, Proc.\ Camb.\ Phil.\ Soc.\  {\bf 31},
555 (1935). It is interesting to provide a full quotation: ``When two
systems, of which we know the states by their respective representation, enter
into a temporary physical interaction due to known forces between them and when
after a time of mutual influence the systems separate again, then they can no
longer be described as before, viz., by endowing each of them with a
representative of its own. I would not call that {\it one} but rather {\it the}
characteristic trait of quantum mechanics." For further discussion, see also
M.  Horne, A.  Shimony and A.  Zeilinger, in {\it Quantum Coherence}, J. S.
Anandan, ed., World Scientific, Singapore (1990).

 \pritem{\mann} A. Mann and M. Revzen, eds., {\it The Dilemma of Einstein,
Podolsky and Rosen---60 Years Later}, Inst.\ Phys.\ Pub., Bristol (1996) (Ann.\
Israel Phys.\ Soc.\ {\bf 12}).

 \pritem{\destruct} L. S. Schulman, Destruction of Interference by
Entanglement, Phys.\ Lett.\ A {\bf 211}, 75 (1996).

 \pritem{\garg} A. Garg, Decoherence in Ion Trap Quantum Computers, Phys.\
Rev.\ Lett.\ {\bf 77}, 964 (1996).

 \pritem{\chuang} Among many recent references, see for example, I. L. Chuang,
R. Laflamme, P. W. Shor and W. H. Zurek, Quantum Computers, Factoring, and
Decoherence, Science {\bf 270} (8 December), 1633 (1995).

 \pritem{\timebook} L. S. Schulman, {\it Time's Arrows and Quantum
Measurement}, Cambridge Univ.\ Press, Cambridge (1997).

 \pritem{\bounds} L. S. Schulman, Bounds on decoherence and error, preprint.

\end